\documentclass[preprintnumbers,nofootinbib,noshowpacs,eqsecnum,prd,twocolumn]{revtex4}

\usepackage{graphicx}
\usepackage{amsmath,amsthm,amssymb,amsfonts}
\usepackage{mathrsfs,bbm,slashed}
\usepackage{color,hyperref}

\definecolor{lgray}{gray}{0.9} 		
\graphicspath{ {figures/} }

\makeatletter       

\renewcommand{\p@subsection}{}

\makeatother

\newcommand*{\eweakgroup}{\mbox{$SU(2)_L \times U(1)_Y$} }

\newcommand*{\trans}{\mathrm{T}}                     

\DeclareMathOperator{\BR}{BR}			

\begin{document}

\title{The two-real-singlet Dark Matter model}

\author{A. Arhrib}
    \email[E-mail: ]{aarhrib@gmail.com}
\affiliation{Abdelmalek Essaadi University, Faculty of Sciences and Techniques,
B.P 416, Tangier, Morocco.}

\author{M. Maniatis}
    \email[E-mail: ]{maniatis8@gmail.com}
\affiliation{Departamento de Ciencias B\'a{}sicas, 
Universidad del B\'i{}o B\'i{}o, Casilla 447, Chill\'a{}n, Chile.}


\begin{abstract}
We revisit a Dark Matter model with an extension of the Standard Model 
with two real singlets $\chi$ and $\eta$ obeying a
$\mathbbm{Z}_2 \otimes \mathbbm{Z}_2'$ symmetry, where 
$\mathbbm{Z}_2'$ is broken spontaneously.
While $\chi$ serves as a stable Dark Matter candidate providing the relic density, the real $\eta$ field 
plays the role of a light mediator. 
We study the viability of this model with respect to Dark Matter self-interactions which
may explain the density profiles observed in dwarf galaxies up to scales of the size of our Milky Way.
Moreover, the Standard Model-like Higgs boson of the model has a tiny mixing with the mediator field and
 appears to be consistent with current LHC data.
In this rather minimal extension of the Standard Model the mediator $\eta$ decays naturally 
into Majorana neutrinos neither disturbing standard big bang nucleosynthesis nor cosmic microwave background observations. 
\end{abstract}


\maketitle


\section{Introduction}
\label{intro}

The simplest extension of the Standard Model in order to accommodate Dark Matter is one additional real scalar $\chi$ \cite{Silveira:1985rk,McDonald:1993ex,Burgess:2000yq}.
For a recent overview of this minimal
extension, in particular with respect to the current status of detection limits, see \cite{Athron:2017kgt,Arcadi:2017kky} and references therein. 
The Dark Matter candidate $\chi$ needs to be stable on cosmological time scales and escape all the direct and indirect detection searches. 
Apart from giving the mass of the halo, as indirectly observed in galaxy rotations, Dark Matter should also 
provide the observed density profile in Dwarf galaxies: simulations of collisionless Dark Matter predict a cusp in the density profile in dwarf galaxies which is not observed as inferred from the rotational curves~\cite{Donato:2009ab}. 
This core-cusp problem can be avoided if Dark Matter has appropriate self interactions. 

Let us mention the recent observation \cite{vanDokkum:2018vup}
of an exceptional case of the NGC1052--DF2 galaxy with a stellar mass of approximately $2 \cdot 10^8$ solar masses and
a rotational movement indicating a negligible Dark halo Mass, that is,
without any indirectly detected Dark Matter. This observation clearly disfavors modifications
of gravity as an alternative to explain the indirect evidences of Dark Matter. \\

Here we would like to revisit a model 
which contains the Standard Model Higgs doublet $\varphi$ extended by 
 a real scalar $\chi$, which will act as a Dark Matter candidate, and 
an additional real singlet $\eta$ \cite{Abada:2011qb, Ahriche:2013vqa} mediating
the self interaction of $\chi$. In particular, the mediator $\eta$ is expected to decay sufficiently 
fast in order not to stay in contradiction to standard big bang nucleosynthesis. We will follow a similar approach as in \cite{Ma:2017xxj}, where the scalar $\eta$ couples to Majorana neutrinos and not to charged leptons which would
disturb the cosmic microwave background observations  \cite{Galli:2009zc,Bringmann:2016din}. We shall show that the two-real-singlet model,
with an appropriate assignment of discrete symmetries, can provide the expected relic abundance of  Dark Matter, accomplish self interactions with a mediator which decays sufficiently fast neither disturbing standard big bang nucleosynthesis nor the measurement of the cosmic microwave background. In a numerical study we shall detect parameters of
the model fulfilling the direct and indirect searches for Dark Matter.
Eventually, from the mixing of the mediator with the ordinary Higgs boson, we will show that the
possible enhancement of invisible Higgs decays at the LHC is negligible.


\section{The two singlet model}
\label{twosinglet}

The potential of the model consisting of one doublet $\varphi$, and two real singlets $\chi$ and $\eta$ is
\cite{Abada:2011qb, Ahriche:2013vqa} 
\begin{multline} \label{pot}
V_{\text{2 singlet}} =
\mu_h^2 \varphi^\dagger \varphi
+ \mu_\chi^2 \chi^2
+ \mu_\eta^2 \eta^2
+ \lambda_h (\varphi^\dagger \varphi)^2
+ \lambda_\chi \chi^4
+ \lambda_\eta \eta^4
\\
+ \lambda_{h \chi} (\varphi^\dagger \varphi) \chi^2
+ \lambda_{h \eta} (\varphi^\dagger \varphi) \eta^2
+ \lambda_{\chi \eta} \chi^2  \eta^2 \;.
\end{multline}
Apart from the Standard Model symmetries we have
an additional symmetry $\mathbbm{Z}_2$ with
$\chi \to -\chi$ as well as $\mathbbm{Z}_2'$ with $\eta \to -\eta$ and all other Standard Model fields 
transforming trivially under these discrete symmetries. 
We assume that the neutral component of the doublet $\varphi$ as well as the singlet $\eta$ get vacuum-expectation values (vev's),
$v_h$ and $v_\eta$, respectively, and in addition to the electroweak group, \eweakgroup, it is supposed that
 $\mathbbm{Z}_2'$ is spontaneously broken.
Since $\mathbbm{Z}_2$ is kept intact, we get in this way a stable particle $\chi$ which becomes
a Dark Matter candidate. On the other hand, trilinear couplings arise from the spontaneously broken 
$\mathbbm{Z}_2'$ which provide self-interacting Dark Matter with $\eta$ working as a light mediator particle.

A spontaneously broken discrete symmetry leads to the formation of domain walls, which spoil nucleosynthesis and the cosmic microwave background observations (see for instance \cite{Zeldovich:1974uw}). A solution to this problem is to have the 
discrete symmetry $\mathbbm{Z}_2'$ broken explicitly. This can be achieved by an additional term in the Lagrangian of the form linear in the field, that is, with a cubic mass parameter. We will in the following assume this additional Lagrangian term implicitly, which is however Planck-mass suppressed.
In this way this additional term becomes only effective at energy scales of the order of the Planck mass und can therefore be neglected here.

We suppose to have right-handed neutrinos in the model which 
transform under $\mathbbm{Z}_2'$ as
$\nu_{R, e} \to +\nu_{R, e}$, $\nu_{R, \mu} \to -\nu_{R, \mu}$, and 
$\nu_{R, \tau} \to -\nu_{R, \tau}$. 
We add to the Lagrangian the Majorana kinetic and mass terms
\begin{equation} \label{majorana}
{\cal L}_{\nu_R} = i \bar{\nu}_{R, i} \slashed{\partial}  \nu_{R, i}
- \left( \frac{\lambda_{i j}}{2}\; \eta \; {\bar{\nu}_{R, i}}^c \nu_{R, j} + h. c. \right) \;,
\end{equation}
where $i, j  = e, \mu ,\tau$ denote the three flavors of the neutrinos. 
We note that also a Dirac term is possible for the right-handed electron neutrino
but for simplicity we assume its coupling to be negligible here. Let us mention that 
different assignments of the right-handed neutrinos with respect to $\mathbbm{Z}_2'$ could 
be made.
From the
charges under $\mathbbm{Z}_2'$ we see that the $3 \times 3$ matrix $(\lambda_{i j})$ in \eqref{majorana}
can only have non-vanishing entries in the first row and column except for the diagonal
entry.

Stability, considering only the quartic terms of the potential \eqref{pot}, yields the requirements 
$\lambda_h, \lambda_\chi, \lambda_\eta \ge 0$, and
$\lambda_h + \lambda_\chi + \lambda_{h \chi} \ge 0$, 
$\lambda_h + \lambda_\eta + \lambda_{h \eta} \ge 0$,
$\lambda_\chi + \lambda_\eta + \lambda_{\chi \eta} \ge 0$,
Since we want to have non-vanishing vev's $v_h$, $v_\eta$, but $\mathbbm{Z}_2$ unbroken,
we find the constraints
\begin{multline}
\mu_h^2, \mu_\eta^2 < 0, \quad \mu_\chi^2 \ge 0, \quad
\lambda_h, \lambda_\eta > 0, \quad 
\lambda_\chi \ge 0, \quad \text{and }
\\
\lambda_h + \lambda_\chi + \lambda_{h \chi}  \ge  0, \quad
\lambda_h + \lambda_\eta + \lambda_{h \eta}  \ge 0, \quad
\lambda_\chi + \lambda_\eta + \lambda_{\chi \eta}  \ge 0.
\end{multline}

We note that by a \eweakgroup gauge transformation we can always achieve the form
$\varphi = \begin{pmatrix} 0 , h \end{pmatrix}^\trans$ with $h$ real. 
The tadpole conditions with $\langle h \rangle \neq 0$, 
 $\langle \eta \rangle \neq 0$, and  $\langle \chi \rangle = 0$ read
 \begin{equation}
 \begin{split}
& \langle h \rangle^2 = \frac{1}{2} v_h^2 = \frac{ \lambda_{h \eta} \mu_\eta^2 - 2\lambda_\eta \mu_h^2}{ 4 \lambda_h \lambda_\eta - \lambda_{h \eta}^2}, \\
&\langle \eta \rangle^2 = \frac{1}{2} v_\eta^2 =  \frac{ \lambda_{h \eta} \mu_h^2 - 2\lambda_h \mu_\eta^2}{ 4 \lambda_h \lambda_\eta - \lambda_{h \eta}^2} . 
\end{split}
\end{equation}
These relations may be used to fix the parameters $\mu_h^2$ and $\mu_\eta^2$ in terms of the
vev's $v_h$ and $v_\eta$. that is, 
\begin{equation}
\mu_h^2 = - (v_h^2 \lambda_h + \frac{1}{2} v_\eta^2 \lambda_{h \eta}), \quad
\mu_\eta^2 = - (v_\eta^2 \lambda_\eta + \frac{1}{2} v_h^2 \lambda_{h \eta}).
\end{equation}
 The two Higgs bosons $h$ and $\eta$ mix and form the states $h'$ and $\eta'$.
 In the basis $( h, \eta)$  the corresponding 
 squared mass matrix reads
 \begin{equation} \label{hetamix}
 \begin{pmatrix}
 2 v_h^2 \lambda_h & 
 v_h v_\eta \lambda_{h \eta} \\
 v_h v_\eta \lambda_{h \eta}  &   2 v_\eta^2 \lambda_\eta
 \end{pmatrix},
 \end{equation}
 which is, as usual, diagonalized by an orthogonal matrix with mixing angle $\theta$ given by
 \begin{equation} \label{theta}
 \tan ( 2 \theta) = 
 \frac{v_h v_\eta \lambda_{h \eta} }
 {v_h^2 \lambda_h -  v_\eta^2 \lambda_\eta}\;.
 \end{equation}
 Since we want to have a light mediator $\eta'$, the mixing angle
 $\theta$ has to be small, that is, the doublet $h'$ 
 has only a small contribution from the singlet $\eta$.
 Note that due to the  $\mathbbm{Z}_2 \otimes  \mathbbm{Z}_2'$ symmetry, the
 Yukawa interactions of $h'$ are not changed compared to the Standard Model but deviations arise from the small mixing with angle given in \eqref{theta}.

From the scalar potential~(\ref{pot}),  the trilinear couplings are given by
 \begin{equation} \label{trilinear}
 \begin{split}
 \mu_{\chi \chi \eta'} = & \sqrt{2} (s_\theta v_h \lambda_{h \chi} 
 + c_\theta  v_\eta \lambda_{\chi \eta}), \\
 \mu_{h' \chi \chi}  = & \sqrt{2} (c_\theta v_h \lambda_{h \chi} 
 - s_\theta  v_\eta \lambda_{\chi \eta}), \\
 \mu_{h' h' h'}  = & \sqrt{2} (2 c_\theta^3 v_h \lambda_{h} 
 - c_\theta^2 s_\theta v_\eta \lambda_{h \eta}
 + c_\theta s_\theta^2 v_h \lambda_{h \eta}
 - 2 s_\theta^3 v_\eta \lambda_{\eta} ),
 \\
 \mu_{\eta' \eta' \eta'} = & \sqrt{2} (2 s_\theta^3 v_h \lambda_{h} 
 + c_\theta s_\theta^2 v_\eta \lambda_{h \eta}
 + c_\theta^2 s_\theta v_h \lambda_{h \eta}
 + 2 c_\theta^3 v_\eta \lambda_{\eta}),
  \\
 \mu_{h' \eta' \eta'}  = & \sqrt{2} (2 c_\theta s_\theta^2 v_h (3 \lambda_{h} - \lambda_{h \eta})
 - s_\theta^3  v_\eta \lambda_{h \eta}
 + c_\theta^3 v_h \lambda_{h \eta}
 \\
 &
 + 2 c_\theta^2 s_\theta v_\eta (\lambda_{h \eta} - 3 \lambda_{\eta})),
  \\
 \mu_{h' h' \eta'}  = & \sqrt{2} (2 c_\theta^2 s_\theta v_h (3 \lambda_{h} - \lambda_{h \eta})
 + c_\theta^3  v_\eta \lambda_{h \eta}
 + s_\theta^3 v_h \lambda_{h \eta}
 \\
 &
 - 2 c_\theta s_\theta^2 v_\eta (\lambda_{h \eta} - 3 \lambda_{\eta})),
 \end{split}
 \end{equation} 
 with $s_\theta = \sin(\theta)$ and $c_\theta = \cos(\theta)$, where the mixing angle $\theta$ is given
 in \eqref{theta}. Let us also give the quartic couplings
 \begin{equation} \label{quartic}
 \begin{split}
 \lambda_{\chi \chi \eta' \eta'} & = s_\theta^2 \lambda_{h \chi} 
 + c_\theta^2 \lambda_{\chi \eta},\\
 \lambda_{h' h' \chi \chi} & = c_\theta^2 \lambda_{h \chi} 
 + s_\theta^2 \lambda_{\chi \eta}, \\
 \lambda_{h' \chi \chi \eta'} & = 
 2 c_\theta s_\theta (\lambda_{h \chi} - \lambda_{\chi \eta}) .
 \end{split}
 \end{equation}
 
 Eventually we consider the right-handed neutrinos. Apart from the Yukawa couplings in
  \eqref{majorana} we get from $\langle \eta \rangle$ spontaneously broken a  
 mass matrix 
 \begin{equation} \label{Mneutrino}
 (M_\nu)_{ij} = \lambda_{ij} \langle \eta \rangle, \quad i,j = e , \mu, \tau \;.
 \end{equation}
 

\section{Dark Matter phenomenology}

With the model set up in the previous section we now study the essential aspects of its
phenomenology. In this way we roughly estimate the range of parameters consistent with
different astronomical and laboratory observations. 

\begin{itemize}

\item Estimate of relic abundance of Dark Matter.

The relic abundance benchmark value for Dark Matter annihilation is
$\sigma \cdot v_{\text{rel}} \approx 3 \cdot10^{-26} \text{cm}^3/s$. In the
model considered here, the Dark Matter annihilation occurs into particles with
even $\mathbbm{Z}_2$ symmetry, that is, 
$\chi \chi \to \eta' \eta' / h' h'$.
Besides quartic couplings we encounter $s$ channel contributions with propagators of $\eta'$ and $h'$
as well as $t$ and $u$ channel contributions with a Dark Matter $\chi$ propagator. 
As a first estimate let us assume that the quartic couplings are dominant, because they do not have any propagator suppression.
We thus get from the quartic couplings roughly
\begin{equation}
\qquad\sigma (\chi \chi \to \eta' \eta' + h' h') 
v_{\text{rel}} \approx 
\frac{
	{\lambda}^2_{\chi \chi \eta' \eta'}\!\! + \!\!{\lambda}^2_{h' h' \chi \chi} \!\!
	+ \!\!{\lambda}^2_{h' \chi \chi \eta'}
	}
	{ 32 \pi m_\chi^2}
\end{equation}
which gives the right value for the relic abundance of Dark Matter 
$\chi$ with $m_{\chi}= 100 \text{ GeV}$ for instance for 
${\lambda}_{\chi \chi \eta' \eta'} =  0.05$ and vanishing
${\lambda}_{h' h' \chi \chi}$  and
${\lambda}_{h' \chi \chi \eta'}$.
In the next section we study the accurate numerical solutions of the corresponding Boltzmann
equation describing the number density of Dark Matter. 

\item Self-interacting Dark Matter.

The benchmark value for self-interacting Dark Matter is
$\sigma(\chi \chi \to \chi \chi)/m_\chi \approx 1 \text{ cm}^2 / \text{g}$. 
In the two-singlet model, this self interaction arises quite naturally
from the mediator $\eta'$. 
As a first estimate, we compute
the cross section for $\chi + \chi \to \chi + \chi$ near threshold and with
the mediator mass $m_{\eta'}$ supposed to be much smaller than the Dark Matter mass $m_\chi$
and find in this limit
\begin{equation}
\sigma (\chi \chi \to \chi \chi)  \approx \frac{\mu_{\chi \chi \eta'}^4}{16 \pi m_\chi^2 m_{\eta'}^4} .
\end{equation}

Let us note that the contributions from the quartic $\chi$ interaction of the 
potential~\eqref{pot} are suppressed by the quartic power of the Dark Matter mass while 
the contribution from s-channel $h'$ exchange is suppressed by $1/m_{h'}^4$ compared to the mediator $\eta'$ contribution.
We get the benchmark value easily,
for instance, for $m_\chi = 100 \text{ GeV}$, 
$m_{\eta'} = 20 \text{ MeV}$, and $\mu_{\chi \chi \eta'} = 13.9 \text{ GeV}$.
We note that in the case of a small $h$-$\eta$ mixing, the trilinear coupling $\mu_{\chi \chi \eta'}$ \eqref{trilinear} as well as the quartic coupling 
$\lambda_{\chi \chi \eta' \eta'}$ \eqref{quartic} are dominated by the $\lambda_{\chi \eta}$ parameter. However we can tune this parameter in order to get the required relic density and 
self-interacting Dark Matter. 

\item Decay of the mediator $\eta'$ into neutrinos.

From the Yukawa interaction in \eqref{majorana} we have a mediator boson $\eta'$ which 
couples to the right-handed neutrinos $\nu_R$. Due to the $ \mathbbm{Z}_2'$
even properties of the charged leptons, they only couple very weakly to $\eta'$, that
is only through its $h$ component in contrast to the neutrinos.
This absence of $\eta$ decays into charged leptons avoids any disturbance of the
cosmic microwave background. Specifically, the life time of $\eta'$ should be 
below the benchmark value of one second.
Indeed, with the mixing matrix \eqref{Mneutrino} we find 
a rather fast decay of $\eta'$  into neutrinos
\begin{equation}
\Gamma_{\eta'} = \frac{m_{\eta'} \sum_{ij} |m_{ij}|^2}{32 \pi v_\eta^2}.
\end{equation}
For instance for $\sum_{ij} |m_{ij}|^2 = 0.1 \text{ eV}^2$, and $m_{\eta'} = 20 \text{ MeV}$,
$v_\eta = 5 \text{ GeV}$ we find for the life time $1/\Gamma_{\eta'} \approx 0.8 \text{ s}$, that is,
 below the required limit of one second. \\
 Note that $\eta'$ could also decay through loops to $\eta' \to gg$ and $\eta' \to \gamma \gamma$.
 The couplings of $\eta'$ to the Standard Model particles in the loop are suppressed by the tiny mixing $\theta$, therefore the decays $\eta' \to gg$ and $\eta' \to \gamma \gamma$  have both, a couplings suppression as well as loop-factor suppression.
From the estimated life time of the mediator we expect that it is in thermal equilibrium until it decays away at 
a temperature of about 1 MeV, much later than freeze out of $\chi$ occurring at a temperature of about $T = m_\chi/x_f \approx 4 \text{ GeV}$.

\item Invisible/non-detected decay of the Standard Model-like Higgs boson.

As we have seen previously, the Standard Model-like Higgs boson $h'$ has a tiny component from the $\eta$ field.
All the Higgs couplings to Standard Model particles (including the one-loop diagrams $h' \to \gamma\gamma$, $h' \to \gamma Z$ 
and $h' \to gg$)  will be modified by an extra factor $\cos(\theta)$ \eqref{theta}
which however is close to unity in our case. Therefore, all the Higgs observables in the model considered here are consistent
 with the LHC measurements.  \\
 However, we also have to consider the invisible/non-detected decays of the Standard Model-like Higgs boson. 
 Both experiments, ATLAS and CMS, have performed searches for such invisible/non-detected Higgs decays 
 and set a limit on its branching fraction 
  \cite{Aaboud:2017bja, Khachatryan:2016whc}. CMS considers a combination of several production channels 
  and gives an upper limit of 24\% on  $\BR (h'\to \text{invisible})$\footnote{Similar results are obtained from global fit analysis of LHC data, but they are model dependent.} at the 95$\%$ confidence level \cite{Khachatryan:2016whc}. 
 
 In the model considered here, $\eta'$ couples to Standard Model particles only through the tiny mixing $\sin(\theta)$. 
 All the production mechanisms of $\eta'$ will be suppressed such that they have escaped both at LEP and at the LHC.  
 The field $h'$ can decay into one of the following invisible/non-detected final states: $h'\to \nu\nu$,  $h'\to \chi\chi$,  
 $h'\to \chi\chi \eta'$ if kinematically allowed, and also $h'\to \eta'\eta'$. The first decay takes place only through the tiny mixing angle $\theta$ and is therefore suppressed.  
  In a scenario with $m_\chi\approx 100$ GeV both $h'\to \chi\chi$ and  $h'\to \chi\chi \eta'$  are kinematically forbidden
  while $h'\to \eta' \eta'$ is accessible and its branching fraction should be smaller than 0.24. 
  In the case of a small mixing angle $\theta$ where $\mu_{h'\eta'\eta'} \approx \sqrt{2} v_h \lambda_{h\eta}$ \eqref{trilinear}, 
  this limit on the branching ratio translates into an upper limit on the coupling, that is, $\lambda_{h\eta} \leq 1.3 \cdot 10^{-2}$, which can be satisfied in the model. 
 
\end{itemize}

\section{Numerical study}
\begin{figure}[htp]
\includegraphics[width=0.5\textwidth]{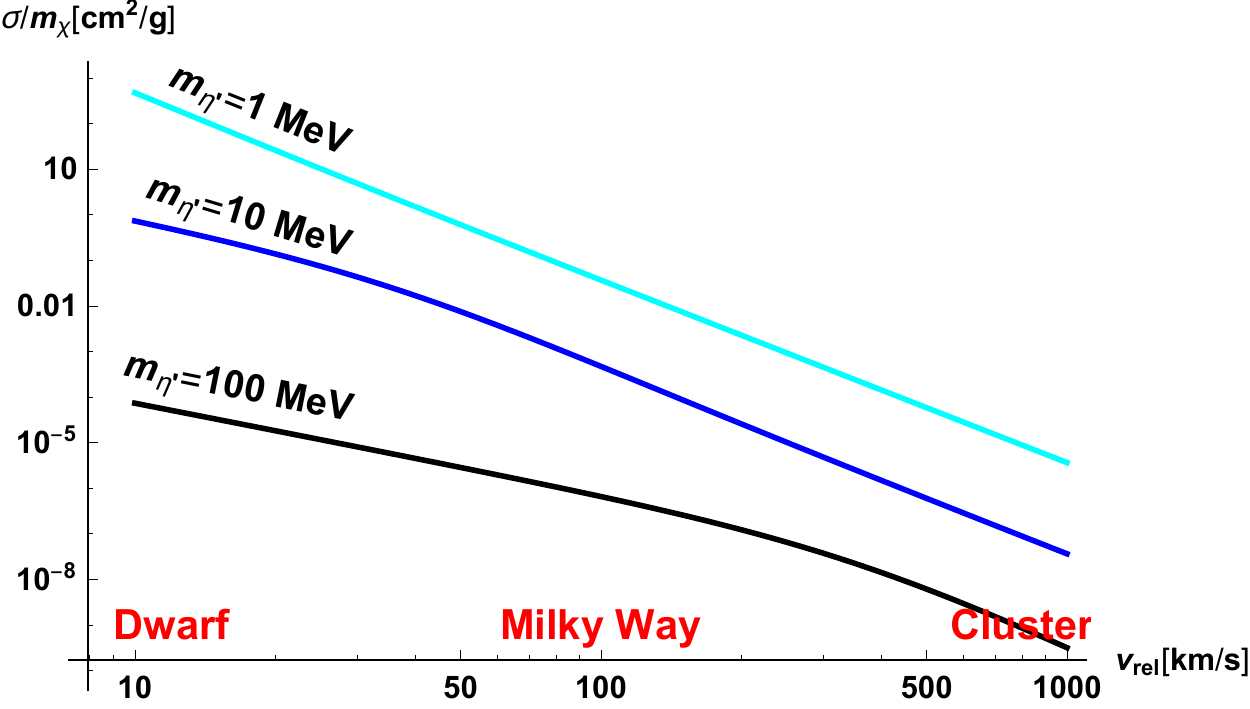}
\caption{\label{crossvel} \em Dependence of the elastic cross section $\sigma(\chi \chi \to \chi \chi)/m_\chi$ on the relative velocity $v_\text{rel}$. 
The relative velocity is of the order of typically 10 km/s for dwarf galaxies, 200 km/s in Milky Way type galaxies and 
1000 km/s in galaxy clusters. For different choices of the mediator particle mass, that is, $m_{\eta'}=1,\; 10,\; 100 \text{ MeV}$, 
the elastic cross section per Dark Matter mass is shown.} 
\end{figure}

In this section we shall present a more detailed calculation of the phenomenology of the two-real-singlet Dark Matter model. 
We study the direct and indirect Dark Matter detection predictions, the 
self-interacting Dark Matter cross sections with respect to different astronomical scales as well 
as the Higgs-boson decays.
We begin with a closer study of self interactions with respect to different astronomical scales. The 
observations show typically a dense core in dwarf galaxies up to galaxies of the size of the Milky Way. From simulations
of collisionless Dark Matter in contrast a cusp in the density profile of smaller galaxies is to be expected.
However,
this core-cusp problem does not arise  
at larger scales of galaxy clusters (see for instance \cite{Tulin:2017ara}).
Whereas the clumping behavior on smaller scales (dwarf and Milky Way type galaxies) can be explained by 
cross sections of self interactions per Dark Matter mass of the order of $\sigma/m_\chi \approx 1 \text{ cm}^2/g$, the absence of 
this clumping at larger scales is consistent with $\sigma/m_\chi < 0.1 \text{ cm}^2/g$. The typical relative velocity in 
a dwarf galaxy is of the order of $10 \text{ km}/s$, in galaxies of the size of our Milky Way it is of the order of $200  \text{ km}/s$, whereas 
at cluster scales this velocity is as large as about $1000  \text{ km}/s$. 
In order to study this velocity dependence in the two-real singlet Dark Matter model considered here, we compute the
elastic cross section $\sigma(\chi \chi \to \chi \chi)$. In the scalar potential \eqref{pot}, the negative parameter $\mu_\eta^2$
provides spontaneous symmetry breaking with respect to the $\eta$ field. 
Therefore, we get in addition to the quartic $\chi$-$\chi$-$\eta'$-$\eta'$ coupling 
 also trilinear $\chi$-$\chi$-$\eta'$ couplings. In the computation of the elastic cross section  
we encounter $s$-, $t$- and $u$-channel diagrams from the trilinear couplings as well as diagrams with a quartic Dark Matter interaction.
In Fig. \ref{crossvel} we present the result for $\sigma(\chi \chi \to \chi \chi)/m_\chi$ depending of the relative
velocity $v_\text{rel}$. We set the vacuum-expectation value to $v_\eta = 5 \text{ GeV}$ in this example and vary the
mass $m_{\eta'}$ as indicated in the figure.  
We see that we can get the required value of at least $\sigma/m_\chi \approx 1 \text{ cm}^2/g$ for small scales in
dwarf and Milky Way type galaxies as well as decreasing
cross sections per Dark Matter mass at cluster scales if we set the mediator mass to $m_{\eta'}  \approx 20 \text{ MeV}$.  \\

In order to determine the relic density of Dark Matter, the corresponding Boltzmann equation which describes the number density
has to be solved. In order to solve this equation numerically
we use the freely available MircOmegas package \cite{Belanger:2013oya} in the version 5.0.4. 
The annihilation
cross sections are computed, which in general are very sensitive to the masses of the particles contributing.
Loop corrections to the mass and to the partial width of the Higgs boson 
are implemented  \cite{Belanger:2013oya}
as well as the Sommerfeld enhancement \cite{Feng:2010zp} of multiple exchanges of mediators, yielding a prediction 
facing the accurate observations from the Planck satellite \cite{Aghanim:2018eyx}.

In addition to the computation of the relic density, MicrOmegas gives the direct detection rates. 
MicrOmegas computes decay rates and cross section via the CalcHEP program package \cite{Belyaev:2012qa}. 
We implement the two-singlet-Dark Matter model in the CalcHEP program package which then in turn is loaded in MicrOmegas. 

In the numerical study, the parameter ranges of the model are chosen as follows:
with view on the self-interacting Dark Matter observations studied above at different
astronomical scales, we vary the mass of the light mediator in the range $20 \text{ MeV} < m_{\eta'} < 50 \text{ MeV}$.
Since the relic density is very sensitive to the Dark Matter mass we restrict the variations to 
$95 \text{ GeV} < m_{\chi} < 110 \text{ GeV}$. For the vacuum expectation value of the $\eta$ field we 
allow a rather large variation, that is, $1 \text{ GeV} < v_{\eta} < 30 \text{ GeV}$. Exclusively varying this
parameter shows a rather slow dependence of the relic density only raising for larger values above about 20 GeV. 
Since we want to have a small mixing of the Higgs boson $h$ with the mediator $\eta$
 we fix the coupling $\lambda_h$ together
with the quadratic parameter $\mu_h^2$ like in the Standard Model, that is, 
$\mu_h = - m_h/\sqrt{2}$, $\lambda_h = -\mu_h^2/v^2$, with $v \approx 246 \text{ GeV}$ and $m_h \approx 125 \text{ GeV}$.
Further, we vary the quartic parameter
$0.1 < \lambda_\chi < 0.17$, keeping it perturbatively small and positive for stability reasons. 
As to be expected, the quartic mixing parameters, $\lambda_{h \chi}$, $\lambda_{h \eta}$, 
$\lambda_{\chi \eta }$ are very sensitive to the relic density of Dark Matter, Higgs decay rates, and 
Dark Matter self-interactions, respectively. We vary each of them separately in order to find approximate appropriate 
values; in detail we choose these couplings in the respective ranges
$0.005 < \lambda_{h \chi}< 0.012$,
$1.1\cdot 10^{-8}< \lambda_{h \eta} < 1.4\cdot 10^{-8}$,
$0.055 < \lambda_{\chi \eta }< 0.065$. 
Note that the  $h$-$\eta$ mixing angle $\theta$ follows from \eqref{theta} depending on the chosen parameters and
turns out to be tiny.
Eventually, the couplings in the Majorana mass terms \eqref{majorana} are chosen
such as to fulfill the condition that the $\eta'$ live time $1/ \Gamma_{\eta'}$ is below one second, not disturbing 
 standard big bang nucleosynthesis.

With these preparations we shuffle the parameters in the mentioned ranges and compute
the relic density from the solution of the Boltzmann equation, the direct Dark Matter-nucleon cross sections,
as well as the branching ratios of the $h$ decay with the help of MicrOmegas.

\begin{figure}[htp]
\includegraphics[width=0.5\textwidth]{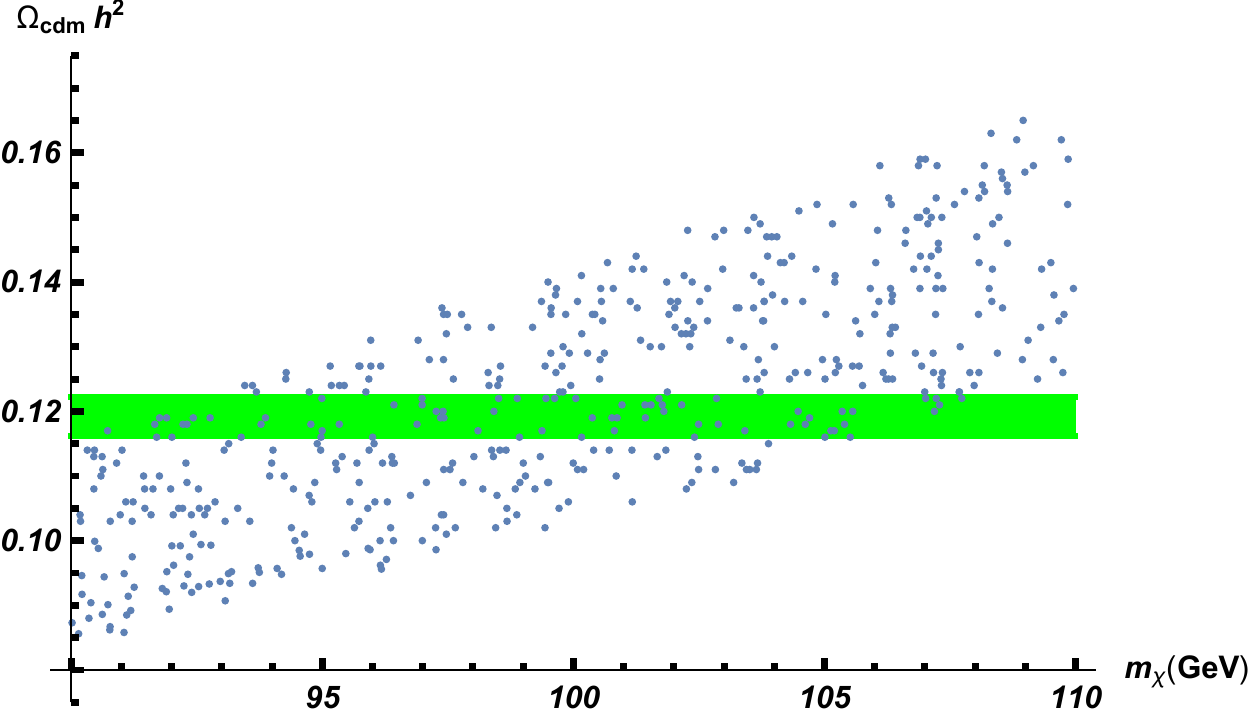}
\caption{\label{rd} \em Scatter plot of the relic density $\Omega_{\text{cdm}} h^2$ of the Dark Matter candidate $\chi$ depending on
the Dark Matter mass $m_{\chi}$. The central green band shows the Planck satellite measurements 
 $\Omega_{\text{cdm}} h^2 =0.1198 \pm 0.0012$ \cite{Aghanim:2018eyx} with three standard deviations.
} 
\end{figure}

In Fig. \ref{rd} we show the  relic densities $\Omega_{\text{cdm}} h^2$ of $\chi$ for all
shuffled parameters depending on the Dark Matter mass $m_\chi$ itself.
We see that the calculated relic density is very sensitive to the Dark Matter mass $m_\chi$. In particular, for a Dark Matter mass in
the range 
$90 < m_\chi <110 \text{ GeV}$ we see that we can get the value observed by Planck \cite{Aghanim:2018eyx}, that is,
$\Omega_{\text{cdm}} h^2 = 0.1198 \pm 0.0012$.  

\begin{figure}[htp]
\includegraphics[width=0.5\textwidth]{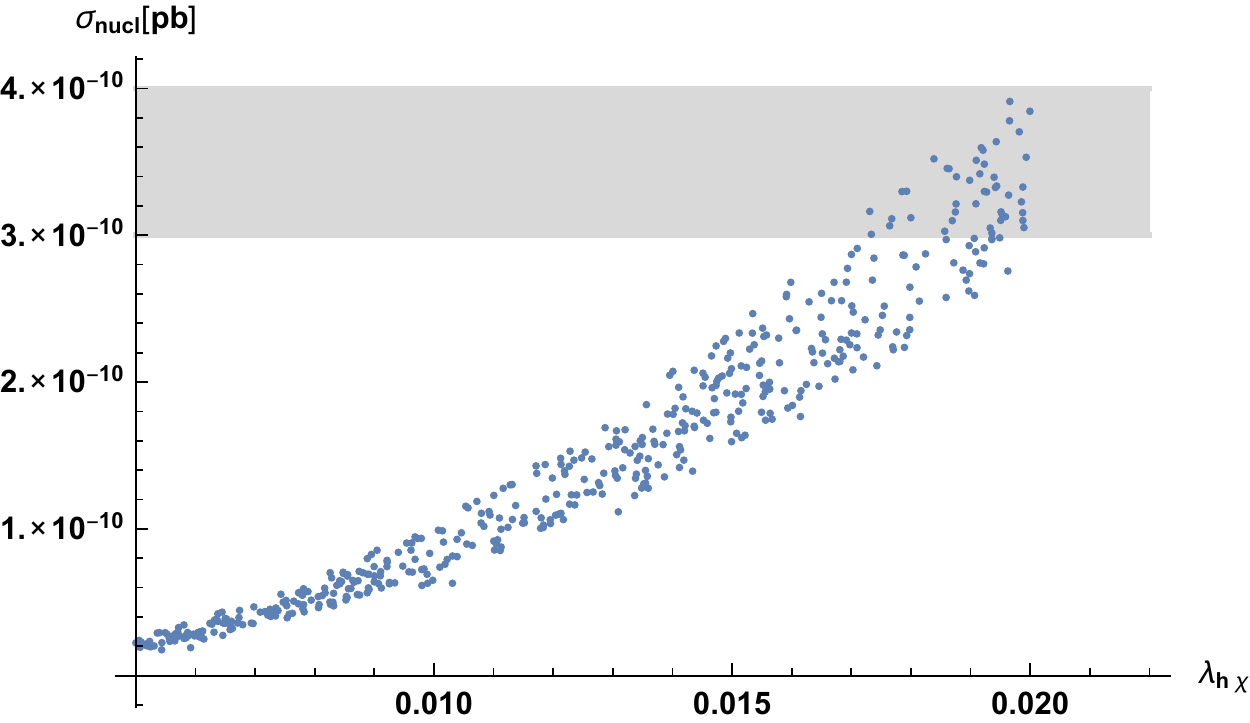}
\caption{\label{sigmaD} \em Scatter plot of the spin-independent direct detection cross section 
$\sigma_{\text{nucl}}$ of Dark Matter-nucleon scattering shown
for different values of the quartic potential parameter $\lambda_{h \chi}$. One of the most stringent limits on 
this cross section comes from the negative searches of the PandaX-II experiment \cite{Cui:2017nnn}.
Their results require for $m_\chi =100 \text{ GeV}$ an upper limit of the
cross sections of about $3\cdot 10^{-10} \text{ pb}$ as shown as shaded region in this plot.
}
\end{figure}

For the shuffled parameters we compute the Dark Matter-nucleon direct detection cross section and 
as to be expected, this cross section is very sensitive to the quartic coupling $\lambda_{h \chi}$. 
This quartic coupling parameter generates trilinear couplings via spontaneous symmetry breaking. 
Therefore, $\lambda_{h \chi}$ provides a coupling of the Dark Matter candidate
via a Higgs boson $h'$ to the baryons of the nucleons.
On the other hand, stringent 
limits on this cross section have been published from direct search experiments, for instance from PandaX-II \cite{Cui:2017nnn}. 
More specifically, their negative search result provides a constraint, that is, for a Dark Matter mass of $m_\chi =100 \text{ GeV}$ 
this constraint corresponds to an upper limit for the spin-independent Dark Matter-nucleon cross section
of about $\sigma_{\text{nucl}} \approx 3\cdot 10^{-10} \text{ pb}$. 
In Fig. \ref{sigmaD} we show the scatter plot of the computed spin-independent Dark Matter-nucleon cross sections,
where the value of the potential parameter $\lambda_{h \chi}$ is given explicitly. 
As we can see in this figure, large parts of the studied parameter space give direct Dark Matter-nucleon cross sections
 below the detection limit.

Eventually, we check the branching ratio of Higgs decays to the mediators, that is, $h' \to \eta' \eta'$. This decay 
channel is expected to be invisible at collider searches like at the LHC due to the dominant subsequent decay of the 
mediators into escaping neutrinos. The trilinear $h'$-$\eta'$-$\eta'$ 
coupling arises from the quartic parameter $\lambda_{h \eta}$ via
spontaneous symmetry breaking. We have chosen rather small values for this parameter in order 
to get a sufficiently large relic density and this in turn 
corresponds to tiny branching ratios of this decay channel of the order of
$BR(\Gamma_{h' \to \eta' \eta'}) \approx 3 \cdot 10^{-13}$, that is, 
far below the invisible detection limits, as measured for instance at the CMS detector at the LHC \cite{Khachatryan:2016whc}.

Altogether we find that with respect to the chosen ranges of parameters, the relic density measurement of
Dark Matter from the Planck satellite  \cite{Aghanim:2018eyx} turns out to provide the most severe constraint.
We find for the fractions of parameter points passing this constraint, that about 4\%, 10\%, 14\% give the right 
relic density with respect to one, two, three standard deviations, respectively. The direct detection limits as shown in
Fig. \ref{sigmaD} only constraint parameter values with a too large mixing parameter $\lambda_{h \chi} > 0.017$. 

To summarize,
we see that the model can be brought into agreement with the direct and indirect detection limits as well as the collider constraints 
for Higgs decays. 

\section{Concluding remarks}
We have revisited the two-singlet extension of the Standard Model  \cite{Abada:2011qb, Ahriche:2013vqa} which accommodates Dark Matter. As usual, the Dark Matter particle
is stabilized by an unbroken $\mathbbm{Z}_2$ symmetry. An additional scalar, odd
under a  spontaneously broken $\mathbbm{Z}_2'$ symmetry, is responsible for the generation
of a mediator particle. 
We have shown that for a rather light mediator mass we get Dark Matter self interactions in agreement with
the observed density profiles of dwarf galaxies and galaxies of the size of our Milky way. 
Besides, the self-interaction cross section has been shown to drop naturally such that
at scales of clusters of galaxies no clumping is to be expected in agreement with observations.

The mediator plays an essential role in this self interactions and in addition,
it decays sufficiently fast and therefore does not disturb
standard big bang nucleosynthesis. Moreover, the mediator decays dominantly into neutrinos; 
therefore there appears no contradiction with the cosmic microwave background observation.
We also have studied the negative results of the direct Dark Matter-nucleon searches, provided
 for instance from the PandaX-II experiment. 
In scatter plots we have shown the results for variations of the parameters of the model and
we have seen, that it is rather straight forward to find agreement with the indirect and direct
Dark Matter searches. The most stringent constraints come thereof from the new indirect observations
of cold Dark Matter by the Planck satellite. 

Since the mediator singlet $\eta$ mixes with the neutral Higgs-boson doublet component $h$, the model predicts deviations in Higgs decays.  This comes from the fact that the singlet component of the mixed state $h'$, \eqref{hetamix}, does not couple to charged leptons in contrast to the SM Higgs boson. An enhancement of an addition decay channel of the $h'$ into pairs of mediators can
in principle spoil the invisible searches at the LHC which agree with the SM. 
However since the corresponding coupling is small enough, the branching ratio
into mediators escapes detection at the LHC.

Eventually
let us summarize that the two-singlet extension is a rather minimal extension of the SM and therefore rather attractive providing a viable Dark Matter candidate.


\begin{acknowledgments}
A.A. acknowledges the Alexander von Humboldt Foundation and the Max Planck Institute for Physics, Munich, where part of this work has been done.
The work of M.M. is supported, in part, by the UBB Grant {\em Cosmolog\'ia y Part\'iculas Elementales}, No. GI 172309/C.
\end{acknowledgments}



\end{document}